\documentclass[twocolumn,showpacs,preprintnumbers,amsmath,amssymb]{revtex4}
\usepackage{graphicx}
\usepackage{dcolumn}
\usepackage{bm}

\begin{document}

\title{Landau-Khalatnikov Circuit model for Ferroelectric Hysteresis}

\author{S. Sivasubramanian}
\author{A. Widom}
\affiliation{Physics Department, Northeastern University, Boston MA 02115}

\date{\today}

\begin{abstract}
We present the circuit equivalent of the Landau-Khalatnikov 
dynamical ferroelectric model. The differential equation 
for hysteretic behavior is subject to numerical computer 
simulations. The size and shape of the simulated hysteretic loops 
depends strongly on the frequency and the amplitude of the driving 
electric field. This dependence makes the experimental 
extraction of the coercive electric field difficult. The 
bifurcation of the driven Landau-Khalatnikov 
model is explained in detail.
\end{abstract}

\maketitle

\section{Introduction}

Many of the interesting properties of ferroelectric materials 
are probed experimentally by measuring the polarization 
response to a time varying electric field. These properties form 
the basis for important potential computer engineering 
applications\cite{1,2,3,4} such as ferroelectric random access 
memory or ferroelectric memory field effect transistors.  
A common method to observe ferroelectric polarization features 
is via the hysteresis loop. A sinusoidal electric field is applied 
to the ferroelectric sample while the polarization 
\begin{math} {\bf P} \end{math} and/or polarization current density 
\begin{math}{\bf J}=(\partial {\bf P}/\partial t) \end{math}  
is continuously monitored. The observed data is then in principle 
utilized to determine the thermal remnant polarization 
\begin{math} {\bf P}_s \end{math} and coercive field 
\begin{math} {\bf E}_c \end{math}. 

Some of the partially understood experimental results are as 
follows: (i) No unique coercive field is evident during the hysteretic 
process of switching. (ii) When an alternating electric field is applied, 
a hysteretic loop is observed only within a limited frequency band 
and a limited range of amplitudes. (iii) Multiple hysteresis loops are 
commonly observed. 

Various approaches have been adopted to explain these strange 
characteristic features of the ferroelectric materials. 
One approach is to create a equivalent circuit 
models\cite{5,6,7,8,9,10,11,12,13} that simulate the experimental 
data. The earliest among these models  
are the Sawyer-Tower circuit\cite{14} and modifications 
thereof\cite{15,16} which are used to study the hysteretic properties 
of  ferroelectric capacitors. Currently, many of the circuit 
models are computed using the SPICE\cite{17} simulations. 
Previous circuit models depend strongly upon which particular  
applications one is interested in analyzing. No single standard circuit 
has been put forward to explain the great variety of experimental results 
available for different regimes of electric field and 
frequency. Our goal is to explore many of the experimentally observed 
properties employing a single circuit model. The model is  
equivalent to the Landau-Khalatnikov dynamical equation for 
the ferroelectric polarization. This non-linear circuit model 
produces a variety of previously unexpected hysteretic 
behaviors.

In Sec.\ II we present the circuit equivalent of the 
Landau-Khalatnikov\cite{18,19,20,21} dynamical ferroelectric model. 
In Sec.\ III the differential equations which must be 
solved are written in the ``dimensionless'' form suitable for 
numerical simulations. In Sec.\ IV the dependence of the size and 
shape of the simulated hysteretic loops on the frequency and amplitude 
of the driving electric field will be exhibited. 
Sec.\ V deals with the bifurcation of hysteretic curves. 
The dynamical properties of the model contains a regime of broken 
symmetry and a regime of unbroken symmetry. The conventional hysteric 
loops are in the symmetric dynamical regime. In the concluding 
Sec.\ VI future experimental prospects for the model are explored.

\section{The Landau-Khalatnikov Circuit}

The thermodynamic equations of state for a ferroelectric model are 
described by the energy per unit volume 
\begin{math} U({\bf P},S) \end{math}
as a function of polarization and entropy per unit volume obeying 
\begin{equation}
dU={\bf E}_{thermal}\cdot d{\bf P}+TdS.
\end{equation}
In addition to the thermal electric field, 
a ferroelectric sample exhibits a dissipative electric field 
derived from an Ohm's law resistivity \begin{math} \rho  \end{math}, 
\begin{equation} 
{\bf E}_{Ohm}=\rho {\bf J}
=\rho\left({\partial {\bf P}\over \partial t}\right).
\end{equation}  
The total electric field  
\begin{math}{\bf E}={\bf E}_{thermal}+{\bf E}_{Ohm}\end{math} 
gives rise to the Landau-Khalatnikov dynamical equation of motion  
\begin{equation}
{\bf E}=\left({\partial U \over \partial{\bf P}}\right)_S+
\rho\left({d{\bf P}\over dt}\right). 
\end{equation} 
Finally, the Maxwell displacement field 
\begin{math} {\bf D} \end{math} within a ferroelectric material is 
given by 
\begin{equation}
{\bf D}=\epsilon_0 {\bf E}+{\bf P}.
\end{equation}

\begin{figure}[bp]
\scalebox {0.5}{\includegraphics{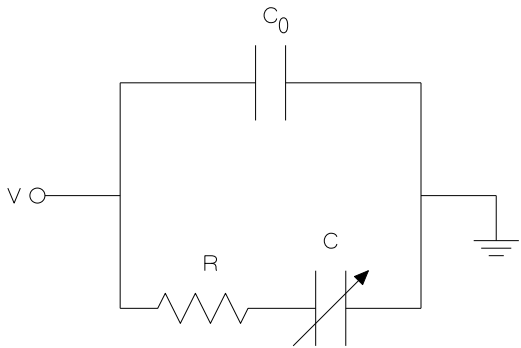}}
\caption{The equivalent circuit to the Landau-Khalatnikov 
dynamical equation is shown in the above figure. 
The linear capacitor is on the upper branch of the circuit 
and the non-linear capacitor is on the lower branch.  
The polarization energy ${\cal U}$ is stored in the 
non-linear capacitor. The ohmic resistor $R$ describes 
the dissipation due to time variations in the 
ferroelectric polarization. The geometric capacitance $C_0$ 
is defined in Eq.(9). $V$ is the applied voltage across 
the parallel circuit.}
\label{fig1}
\end{figure}

If the ferroelectric material is placed inside of a capacitor, 
then the total charge on one capacitor electrode is given 
by the surface integral 
\begin{equation}
{\cal Q}_{tot}=\oint_{electrode} {\bf D}\cdot d{\bf \Sigma }.
\end{equation}
Were the capacitor electrodes embedded in the vacuum, then 
the capacitor charge would be  
\begin{equation}
{\cal Q}_{vac}=
\epsilon_0 \oint_{electrode} {\bf E}\cdot d{\bf \Sigma }.
\end{equation}
In general, Eqs.(4), (5) and (6) imply 
\begin{equation}
{\cal Q}_{tot}={\cal Q}_{vac}+{\cal Q}
\end{equation} 
where 
\begin{equation}
{\cal Q}=\oint_{electrode} {\bf P}\cdot d{\bf \Sigma }.
\end{equation} 

Eq.(7) describes the charges on two capacitors in parallel with the 
geometrical capacitance \begin{math} C_0  \end{math} relates 
the charge to voltage ratio of the vacuum 
\begin{equation}
{\cal Q}_{vac}=C_0 V.
\end{equation}
Eqs.(3) and (8) describe the Landau-Khalatnikov dynamics for 
the non-linear capacitor in FIG.\ 1 with voltage 
\begin{math} {\cal V}({\cal Q}) \end{math} connected in series with 
an Ohm's law resistor \begin{math} R \end{math}; i.e.   
\begin{equation}
V={\cal V}({\cal Q})+R\left({d{\cal Q}\over dt}\right).
\end{equation}
If \begin{math} {\cal U}({\cal Q},{\cal S}) \end{math} denotes 
the energy of the non-linear capacitance, then 
\begin{equation}
d{\cal U}=Td{\cal S}+{\cal V}d{\cal Q}. 
\end{equation}

The circuit version of Eq.(3) is that 
\begin{equation}
V=\left({\partial {\cal U}\over \partial {\cal Q}}\right)_{\cal S}
+R\left({d{\cal Q}\over dt}\right).
\end{equation}
The circuit Eq.(12) is pictured in FIG.\ I. 
In the Landau-Khalatnikov circuit, the upper 
capacitor carries a charge 
\begin{math} {\cal Q}_{vac}=C_0 V \end{math} 
while the lower non-linear capacitor carries a charge 
\begin{math} {\cal Q} \end{math} and voltage 
\begin{math} {\cal V} \end{math} determined by the 
thermodynamic Eq.(11). The resistance 
\begin{math} R  \end{math} describes the dissipation 
present when the polarization varies with time.

\section{Periodic Voltage Sources}

\begin{figure}[bp]
\scalebox {0.5}{\includegraphics{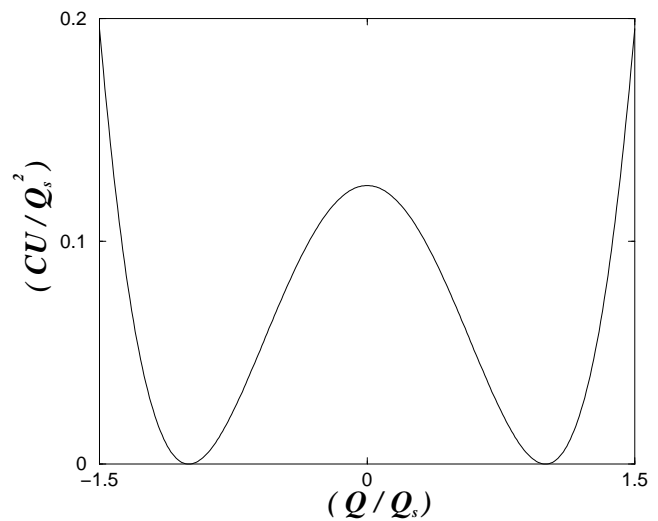}}
\caption{Shown is the Landau energy for a ferroelectric model. 
The energy stored in the non-linear capacitor 
is given by $U=(Q_s^2/8C)\{1-(Q/Q_s)^2)\}^2$ .}
\label{fig2}
\end{figure}

Hysteretic cycles are measured applying a time varying voltage of 
the form 
\begin{equation}
V(t)=V_0 \cos(\omega t).
\end{equation}
For the case of the Landau energy shown in FIG.\ 2,
\begin{equation}
{\cal U}({\cal Q})=\left({{\cal Q}_s^2 \over 8C}\right)
{\left(1-{\left({\cal Q}\over {\cal Q}_s\right)}^2\right)}^2,
\end{equation}
where the saturation polarization \begin{math} {\bf P}_s \end{math}
determines 
\begin{equation}
{\cal Q}_s=\oint_{electrode}{\bf P}_s \cdot d{\bf \Sigma },
\end{equation}
Eqs.(12), (13) and (14) read 
\begin{equation}
R\left( {d{\cal Q}\over dt}\right)+
\left({{\cal Q}\over 2C}\right)
\left\{\left({{\cal Q}\over {\cal Q}_s}\right)^2-1\right\}
=V_0 \cos(\omega t).   
\end{equation} 

In order to solve Eq.(16) numerically, let us introduce 
the dimensionless quantities 
$$
\theta= \omega t, \ \ y=({\cal Q}/{\cal Q}_s), \ \  
\eta =(2\omega RC)^{-1}
$$
and
\begin{equation} 
z=(V_0/R \omega {\cal Q}_s).
\end{equation} 
Eqs.(16) and (17) now read 
\begin{equation}  
\left({dy \over d{\theta}} \right)+ \eta y(y^2-1)=z \cos \theta
\end{equation} 
One seeks solutions to Eq.(18) which are periodic 
\begin{equation}
y(\theta +2\pi ;z,\eta )=y(\theta ;z, \eta ).
\end{equation}
For example, let us suppose that at an initial time zero  
\begin{equation}
y(\theta =0;z,\eta )=x.
\end{equation}
After a numerical integration through one period of motion, 
one then finds that 
\begin{equation}
y(\theta =2\pi ;z, \eta )={\cal G}(x;z,\eta ).
\end{equation}
Eqs.(18), (20) and (21) define the function 
\begin{math} {\cal G}(x;z, \eta )  \end{math}.
It is possible to analytically compute 
\begin{math}  {\cal G}(x;z, \eta )  \end{math} in two limits:
\begin{equation}
\lim_{\eta \to 0} {\cal G}(x;z, \eta )=x,
\end{equation} 
\begin{equation}
\lim_{z \to 0} {\cal G}(x;z, \eta )=
\left({x\ \exp(2\pi \eta )\over 
\sqrt{1+x^2(\exp(4\pi \eta)-1)}}\right).
\end{equation}

A sufficient condition for existence of  
periodic Eq.(19) solutions of Eq.(18) is that 
\begin{math} x  \end{math} be a fixed point (solution) 
of the equation 
\begin{equation}
x={\cal G}(x;z,\eta ).
\end{equation}
If Eq.(24) has a unique solution for  \begin{math} x \end{math}, 
than there exists a unique periodic solution of Eq.(18). 
If Eq.(24) had more than one solution for \begin{math} x \end{math}, 
then there will be (in general) more than one periodic solution 
to Eq.(18). The number of solutions for \begin{math} x \end{math} 
depends on the values of parameters in the 
\begin{math} (z, \eta ) \end{math} plane. 

\section{Hysteresis Curves}

\begin{figure}[bp]
\scalebox {0.5}{\includegraphics{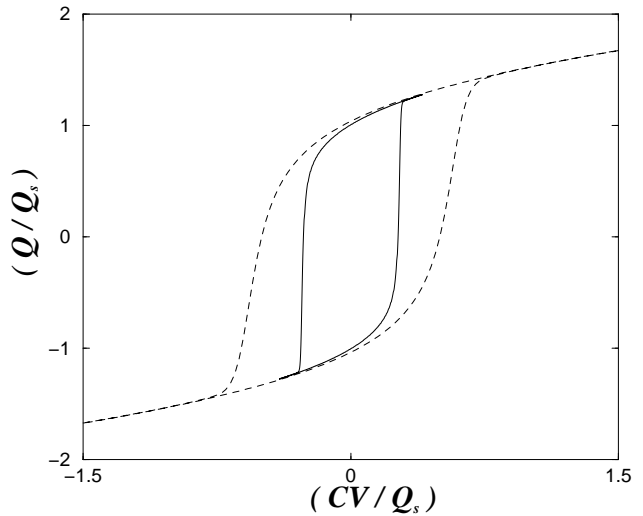}}
\caption{Shown are two different hysteretic loops simulated for 
identical frequencies but differing amplitudes of the 
applied voltage. The smaller inner hysteretic loop corresponds to 
$z=20.0$ and $\eta=25.0$. The larger outer  
hysteretic loop corresponds to $z=200.0$ and $\eta=25.0$. 
Large applied amplitudes imply high apparent values for the 
experimental coercive voltages.}
\label{fig3}
\end{figure}

\begin{figure}[bp]
\scalebox {0.5}{\includegraphics{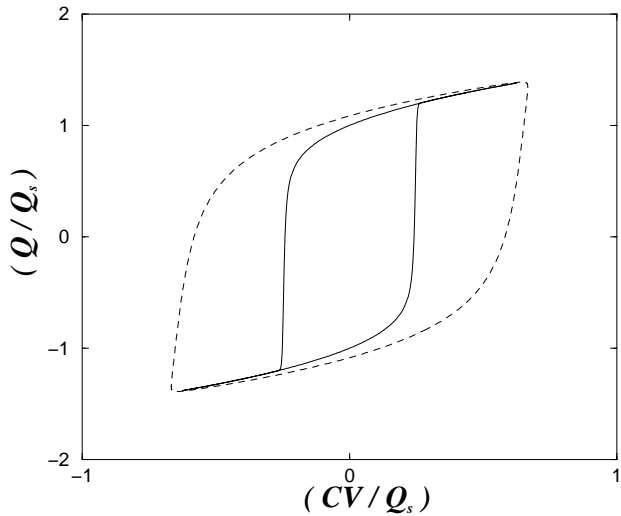}}
\caption{
Shown are two different hysteretic loops simulated for 
identical amplitudes  but differing frequencies of the 
applied voltage. The smaller inner hysteretic loop corresponds 
to $z=100.0$ and $\eta=80.0$. The larger outer  
hysteretic loop corresponds to $z=3.0$ and $\eta=2.25$. 
Large applied frequencies imply high apparent values for the 
experimental coercive voltages.}
\label{fig4}
\end{figure}

In this section we consider the region in the 
\begin{math}(z,\eta ) \end{math} plane for which 
the following conditions hold true: 
(i) Eq.(24) has a unique solution for 
\begin{math} x \end{math} and (ii) there there exists 
a unique periodic solution for 
\begin{math} y(\theta ;z,\eta ) \end{math}. A hysteretic loop 
located in the  \begin{math} (Q,V) \end{math} plane may be found 
by eliminating the time parameter \begin{math} t  \end{math} 
in a one period interval 
\begin{math} 0\le t \le (2\pi /\omega )\end{math} from  
the parametric equations 
\begin{equation}
{\cal Q}={\cal Q}_s y(\theta =\omega t ;z,\eta )\ \ {\rm and}
\ \ V=R\omega {\cal Q}_s z\cos (\omega t).
\end{equation} 

In FIG.\ 3, we have plotted two hysteretic loops corresponding to 
the same frequency but with different driving voltage amplitudes. 
The inner smaller hysteretic loop corresponds to a smaller 
driving voltage than that of the outer larger hysteretic loop.
In FIG.\ 4 we have plotted two hysteretic loops 
corresponding to the same voltage amplitude but with different  
frequencies. The inner smaller hysteretic loop corresponds to a lower  
frequency than that of the outer larger hysteretic loop.

Since the large amplitude and high frequency loops have a larger 
enclosed area than do the small amplitude and low frequency loops,  
it is not at once evident how to extract the coercive voltage 
directly from hysteretic loop data. This is a well known problem 
in attempting to measure the coercive electric field forcing 
a polarization flip. The nature of the problem becomes evident 
from a study of the effective circuit in FIG.\ 1. The 
{\em theoretical} ``coercive voltage'' 
\begin{math} {\cal V}_c \end{math} 
is that voltage across the lower branch non-linear capacitor 
present at the time when the polarization flips. The 
{\em experimental} ``coercive voltage'' 
\begin{math} V^{(exp)}_c=RI_c+ {\cal V}_c \end{math} 
includes the voltage across
the resistor. It is not an easy matter to separate the dissipative 
voltage from the thermal capacitor voltage. 

In order to understand the ``area'' enclosed by the hysteresis loop 
in the  \begin{math} (Q,V) \end{math} plane, one notes that the 
thermodynamic work performed by the non-linear capacitor during 
one cycle is given by  
\begin{equation}
-W({\rm cycle})=\oint_{loop}Vd{\cal Q},
\end{equation}
which is also (via the first law of thermodynamics) the heat 
dissipated by the resistor \begin{math} R \end{math} during 
the cycle. The area enclosed by the hysteretic loop is large 
or small, respectively, when the dissipated heat in the resistor 
is large or small. 

In order to understand the role of dissipative heating in 
more detail, multiply Eq.(10) by the current 
\begin{equation}
I=\left({d{\cal Q}\over dt}\right) 
\end{equation}
and integrate the result over time for one cycle; i.e. 
\begin{equation}
\oint_{loop}VIdt=\oint_{loop}
\left({d {\cal U}\over d {\cal Q}}\right)
\left({d{\cal Q}\over dt }\right)dt
+\oint_{loop}RI^2dt.
\end{equation}
The first term on the right hand side of Eq.(28) obeys 
\begin{equation} 
\oint_{loop} d{\cal U}=0.
\end{equation} 
We then see directly that the work done by the non-linear 
capacitor during one cycle must be dissipated as heat 
in the resistor; i.e. Eqs.(26), (27), (28) and (29) imply 
\begin{equation}
-W({\rm cycle})=\oint_{loop}VIdt=\oint_{loop}RI^2dt.
\end{equation} 
Thus, a large applied voltage amplitude 
and/or a large applied voltage frequency yield a 
large area loop because more heat is dissipated 
per cycle by the resistor.

\section{Dynamical Broken Symmetry}

\begin{figure}[bp]
\scalebox {0.5}{\includegraphics{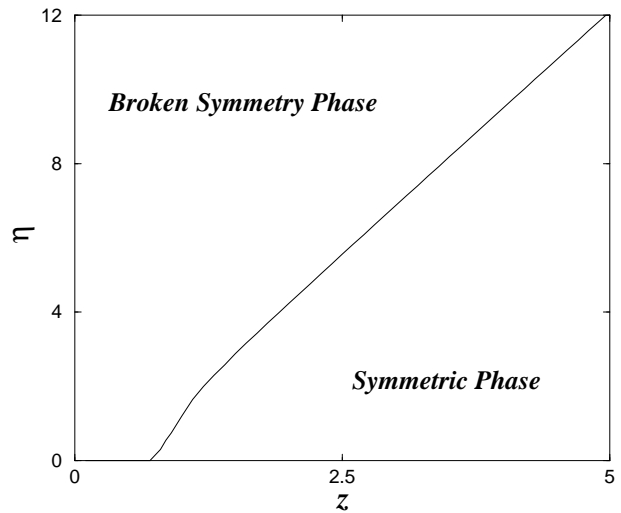}}
\caption{Shown is the phase plane for the 
Landau-Khalatnikov model with the 
dynamical bifurcation curve $z=B(\eta )$. 
If $z<B(\eta )$, then parity symmetry is broken. 
If $z>B(\eta )$, then the parity symmetry is restored.}
\label{fig5}
\end{figure}

\begin{figure}[bp]
\scalebox {0.4}{\includegraphics{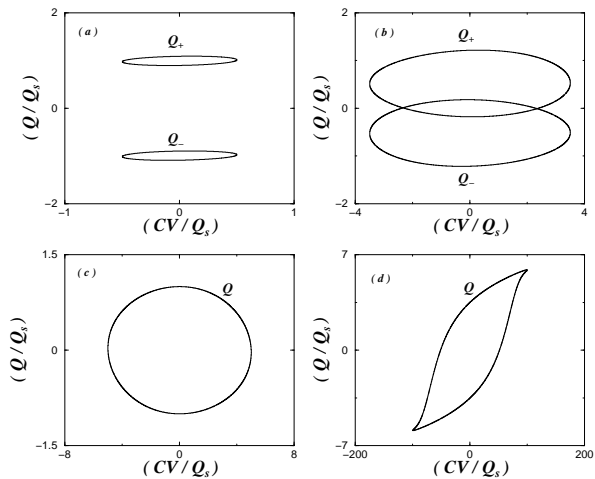}}
\caption{Shown are four loop plots choosing $\eta =0.1$. 
(a) For the small amplitude $B(\eta=0.1)>z_a=0.1$, 
there are two possible small oval 
loops. Only one will be realized in a particular experiment. 
(b) As the voltage amplitude is increased to 
$B(\eta=0.1)>z_b=0.7>z_a$ the ovals grow large but are still 
representative of two different loops. (c) In the symmetric phase 
$z_c=1.0>B(\eta =0.1)$, only a single loop is allowed which does 
not yet look very close to conventional hysteretic loops. 
(d) For large voltage amplitude $z_d=20.0>z_c>B(\eta =0.1)$, 
the conventional hysteretic loop is recovered for the model.}
\label{fig6}
\end{figure}

From the parity symmetry of the energy equation of state 
\begin{math} {\cal U}(-{\cal Q},{\cal S})=
{\cal U}({\cal Q},{\cal S})\end{math}, the are two possible 
thermal {\em equilibrium} minimum energy states at  
\begin{math} {\cal Q}_\pm =\pm {\cal Q}_s \end{math} 
in the absence of a driving voltage. For a {\em small   
driving amplitude} \begin{math} V_0  \end{math} in Eq.(13), 
we then expect two possible periodic charge response functions 
\begin{math} 
{\cal Q}_\pm \left(t+(2\pi/\omega \right)={\cal Q}_\pm (t);   
\end{math}
\begin{equation}
{\cal Q}_+ (t)\ne {\cal Q}_-(t) 
\ \Longrightarrow \ {\rm Broken\ Symmetry}\ .
\end{equation} 
On the other hand, for a {\em large driving amplitude} 
\begin{math} V_0  \end{math} in Eq.(13), 
we expect a unique periodic charge response functions 
\begin{math} 
{\cal Q}\left(t+(2\pi/\omega \right)={\cal Q}(t);   
\end{math}
\begin{equation}
{\cal Q}_+ (t)={\cal Q}_-(t) 
\ \Longrightarrow \ {\rm Restored\ Symmetry}\ .
\end{equation} 

In terms of the dimensionless variables in Eq.(17),  
\begin{equation}
z=V_0/(R\omega {\cal Q}_s)\ \ {\rm and}
\ \ \eta =1/(2\omega RC),
\end{equation}
one expects the two regimes to be described by a 
dynamical bifurcation function \begin{math} B(\eta ) \end{math}
\begin{equation}
z<B(\eta )
\ \Longrightarrow \ {\rm Broken\ Symmetry}\ ,
\end{equation}
\begin{equation}
z>B(\eta )
\ \Longrightarrow \ {\rm Restored\ Symmetry}\ .
\end{equation}
We have numerically computed 
\begin{math} B(\eta ) \end{math}
and the results are plotted in FIG.\ 5. The regions in the 
\begin{math} (z,\eta ) \end{math} plane are exhibited 
corresponding to the symmetric and broken symmetry 
dynamical phases. In the broken symmetry phase, there are two 
possible hysteretic curves; \begin{math} {\cal Q}_+(t)  \end{math} 
is localized in the neighborhood of 
\begin{math} +{\cal Q}_s  \end{math} and 
 \begin{math} {\cal Q}_-(t)  \end{math} is localized in the 
neighborhood of \begin{math} -{\cal Q}_s  \end{math}.

For a more clear picture of the bifurcation\cite{22,23,24} 
boundary curve at \begin{math} z=B(\eta ) \end{math} we show a 
sequence of hysteretic loops in FIG.\ 6. For 
\begin{math} \eta =0.10 \end{math}, we have chosen four 
values for the amplitudes \begin{math} z  \end{math}. 
Explicitly 
\begin{eqnarray}
0<z_a=0.1<z_b=0.7<B(\eta=0.1) \nonumber 
\end{eqnarray} 
for illustrations of broken symmetry, and
\begin{equation}
B(\eta=0.1)<z_c=1.0<z_d=20.0  \nonumber 
\end{equation}
for illustrations of unbroken symmetry. 
The two ovals in both FIG.\ 6(a) and FIG.\ 6(b) correspond, 
respectively, to \begin{math} {\cal Q}_\pm (t) \end{math}
for the values \begin{math} z_a  \end{math} and 
\begin{math} z_b  \end{math}. The symmetry is broken because 
only {\em one} of the ovals (\begin{math} {\cal Q}_+ \end{math} 
or \begin{math} {\cal Q}_-  \end{math} but not both)
will appear in an  experiment described by the model. 
The oval in the unbroken symmetry phase corresponding to 
\begin{math} z_c  \end{math} is unique but does not
appear very similar to conventional hysteretic curves. 
By increasing the amplitude to 
\begin{math} z_d>z_c  \end{math} 
a conventional looking hysteretic loop appears in FIG.\ 6(d).

\section{Conclusions}

It has been shown that the Landau-Khalatnikov model of 
ferroelectric hysteresis has a very rich dynamical structure. 
Only a small part of the symmetric dynamical phase had been 
reported in standard literature. The conventional looking 
hysteretic curves are present theoretically and experimentally 
only in the regime of high driving amplitudes.
The region in the neighborhood of the bifurcation into the broken 
symmetry phase has received little or no experimental attention. 
The numerical simulations of this work indicate what is to be 
expected. Experimental measurements of the dynamical bifurcation 
curve \begin{math} z=B(\eta )  \end{math} of ferroelectrics 
would be of great interest in checking the validity of the 
Landau-Khalatnikov model.

Further insights on how to measure the coercive voltage can be 
obtained by driving the circuit in FIG.\ 1 with a voltage source 
which has both DC and AC components 
\begin{equation}
V(t)=V_{ext}+V_1 \cos (\omega t).
\end{equation}
The DC voltage component \begin{math} V_{ext} \end{math} can be 
put into the energy function via 
\begin{equation}
{\cal U}({\cal Q},{\cal S})\to 
{\cal U}({\cal Q},{\cal S};V_{ext})=
{\cal U}({\cal Q},{\cal S})-V_{ext}{\cal Q},
\end{equation}
leaving Eq.(12) in the form 
\begin{equation}
V_0\cos(\omega t)=
\left({\partial {\cal U}({\cal Q},{\cal S};V_{ext})
\over \partial {\cal Q}}\right)
+R\left({d{\cal Q}\over dt}\right).
\end{equation}

The advantage of employing a DC component is now evident. 
The symmetry breaking of the energy in FIG.\ 2 will now 
be controlled by varying the DC voltage. The DC offset 
will allow for precise measurements of the coercive field.

\vfill 
\eject

\end{document}